\documentclass{camera}

\begin{document}

%
\title{Resummed Transverse Momentum Distribution in $b\rightarrow s\gamma$}

%
\author{R.Sghedoni}

%
\organization{Universit\'a di Parma \& INFN, Parma, Italy}

\maketitle

%
\abstract

We present the complete calculation of the transverse momentum
distribution for the decay $b\rightarrow s\gamma$. The calculation
of the coefficients of the infrared logarithms is performed using
general properties of QCD. The resummation of large infrared
logarithms is accomplished according to usual techniques of
resummation. Non logarithmic terms (constants and function
vanishing at the border of the phase space) are calculated for the
leading operator of the effective hamiltonian, $\hat{O}_7$.

\section{Transverse Momentum Distribution in $b\rightarrow s\gamma$.}
The process $b\rightarrow s \gamma$ is a rare decay of the \it
beauty \rm quark, having inclusive branching ratio \cite{jessop}:
\begin{equation}
BR(b\rightarrow s\gamma)=(3.34\pm 0.30)10^{-4}
\end{equation}
It shows a non trivial interplay of electroweak and strong
interactions: the partonic transition is mediated by a loop even
at the lowest level and an effective hamiltonian
\begin{equation}\label{hamiltoniana}
{\cal H}_{eff} = {G_F \over \sqrt{2}} \ V^*_{ts}V_{tb} \
\sum_{j=1}^8 C_j(\mu) \ \hat{\cal O}_j(\mu)
\end{equation}
is built to integrate out the heavy particles in the loop (a $W$
boson and the $top$ quark)\cite{gsw2}. Strong corrections to the
coefficients $C_j(\mu)$ in the hamiltonian produce logarithmic
terms of the form
\begin{equation}
\alpha_S^n \ \log^n\frac{m^2_W}{m^2_b}
\end{equation}
which can be resummed with renormalization group techniques
\cite{misiak}.
\\
It turns out that the leading operator of the effective
hamiltonian is the so-called magnetic penguin
\begin{equation}
\hat{\cal O}_7 = {e\over 16\pi^2}m_{b,\overline{MS}}(\mu_b)
\overline{s}_{L,\alpha}\sigma^{\mu\nu}b_{R,\alpha}F_{\mu\nu}.
\end{equation}
We have considered the transverse momentum distribution of the $s$
quark with respect to the photon direction \cite{noi}:
\begin{equation}
x=\frac{p^2_\perp}{m^2_b}.
\end{equation}
At the lowest order they are emitted along the same direction
because of momentum conservation, while at higher orders of the
perturbation theory a transverse momentum is generated because of
gluon emissions from the heavy quark ($b$) or the light quark
($s$).
\\
The partially integrated rate
\begin{equation}
D(x)= \int_0^1 \ \frac{1}{\Gamma_0} \ \frac{d\Gamma}{dx^\prime} \
dx^\prime
\end{equation}
at the order $\alpha_S$ can be written as
\begin{equation}
D(x)= 1+ \alpha_S \left[- \frac{A_1}{4}\log^2 x+B_1 \log x + \rm
non ~ log ~ terms \right].
\end{equation}
The logarithms become large near the border of the phase space,
for the emission of soft or collinear (to the emitting quark)
gluons. These logarithms need to be resummed to every order of the
perturbation theory according to well known tools of resummation.
\\
It is well known \cite{cttw} that the resummation of large
logarithms is accomplished by an expression of the form:
\begin{equation}\label{master}
D(x) = K(\alpha_S) \Sigma(x;\alpha_S) + R(x;\alpha_S),
\end{equation}
where
\begin{itemize}
\item
$\Sigma(x;\alpha_S)$ is a universal, process-independent function
resumming the infrared logarithms in exponentiated form.  It can
be expanded in a series of functions as:
\begin{equation}\label{g}
\log \Sigma(x;\alpha_S)= L g_1(\alpha_S L) + g_2 (\alpha_S L)+
\alpha_S g_3(\alpha_S L)+\dots \ ,
\end{equation}
where $L=\log x$ (in general $L$ is a large infrared logarithm).
The functions $g_i$ resum logarithms of the same size: in
particular $g_1$ resums leading logarithms of the form
$\alpha_S^n~ L^{n+1}$ and $g_2$ the next-to-leading ones
$\alpha_S^n ~L^n$;
\item
$K(\alpha_S)$ is a short-distance coefficient function, a
process-dependent function, which can be calculated in
perturbation theory:
\begin{equation}
K(\alpha_S) = 1+\frac{\alpha_S C_F}{\pi} k_1 +
O\left(\alpha_S^2\right).
\end{equation}
\item
$R(x;\alpha_S)$ is the remainder function and satisfies the
condition
\begin{equation}\label{remainder}
R(x;\alpha_S) \rightarrow 0 \ \ \ {\rm{for}} \ \ \  x \rightarrow
0.
\end{equation}
It is process dependent, takes into account hard contributions and
is calculable as an ordinary $\alpha_S$ expansion:
\begin{equation}
R(x;\alpha_S)=\frac{\alpha_S C_F}{\pi} r_1(x)+
O\left(\alpha_S^2\right).
\end{equation}
\end{itemize}
In \cite{noi} the resummation of infrared logarithms has been
performed with next-to-leading accuracy in the function
$\Sigma(x;\alpha_S)$. The coefficients $A_1$ and $B_1$ have been
calculated from general properties of QCD radiation, applying the
perturbative evolution for the light quark, described by the
Altarelli-Parisi kernel, and the eikonal approximation for the
massive quark. The resummation of the logarithms has been
performed in the impact parameter space, taking into account
running coupling effects, and the function $g_1$ and $g_2$ have
been obtained. The occurrence of singularities in the resummed
formula signals that even this improved perturbative approach is
not reliable near the border of the phase space, where non
perturbative effects (related to the initial and final states)
make their appearance. In \cite{noi} the impossibility of defining
a \it shape function \rm \cite{shape} is discussed.
\\
In \cite{noi2} the next-to-leading calculation has been completed
calculating the correction of order $\alpha_S$ to the coefficient
function $K(\alpha_S)$; the remainder function $R(\alpha_S;x)$ has
been also computed introducing harmonic polylogarithms as in
\cite{rv}. In opposition to $\Sigma(x;\alpha_S)$, these two
functions are process dependent and have to be evaluated by the
explicit calculation of one loop Feynman diagrams. This completes
the calculation at the next-to-leading order of the transverse
momentum distribution.
\\
An accurate comparison with experimental data will be interesting
to conclude whether the resummation of large logarithms,
considered a standard tool in $b$ physics, is reliable at low
energies. In fact the resummation of large logarithms has been
developed in the past years to describe processes at higher
energies, for example at LEP, at $Q^2=M^2_{Z^0}$ energies, where
the strong coupling constant $\alpha_S$ is smaller,
$\alpha_S(M^2_{Z^0})\sim 0.1$, and non perturbative effects are
less relevant.
\\
Perturbation theory has been applied to describe decays of the
beauty quark since its discovery. While the expansion parameter
$$\alpha_S(m_B)\sim 0.21$$ being reasonably small, allows one to
trust the computations, it is difficult to directly compare the
perturbative approach with the experimental data. As it is well
known, decay rates do not make good quantities to be compared with
the data, because they are proportional to the fifth power of the
beauty quark mass, a poorly known parameter and because they
involve in principle unknown CKM matrix elements such as $V_{cb},~
V_{ub},~V_{ts},$ etc\dots This means that it is not clear if
discrepancies between theory and data have to be ascribed to a
inaccurate theory or to uncertainties of the parameters.
\\
From this point of view semi-inclusive quantities, such as the
transverse momentum distribution we are considering, can give
information about the reliability of the perturbation theory at
low energies, and can suggest new tools to deal with non
perturbative effects (as in the case of the \it shape function \rm
for the threshold distribution \cite{shape}).
\\
\\
The contents of this paper arise from my work with Prof.
L.Trentadue and Dr. U.Aglietti: I wish to thank Prof. Trentadue
for his collaboration and Dr. Aglietti for his essential help.
Furthermore I would like to thank my proof-reader Franco M. Neri.


\thebibliography{99}

\bibitem{jessop} C.Jessop, SLAC-PUB-9610

\bibitem{gsw2} B.Grinstein, R.Springer, M.Wise, Nucl. Phys. B339, 269
(1990)

\bibitem{misiak} K.Chetyrkin, M.Misiak, M.Munz, Phys. Lett. B400, 206 (1997),
Erratum Phys. Lett. B425, 414 (1998)

\bibitem{noi}U.Aglietti, R.Sghedoni, L.Trentadue, Phys. Lett. B522, 83 (2001)

\bibitem{cttw} S.Catani, L.Trentadue, G.Turnock, B.Webber, Nucl. Phys. B407, 3 (1993)

\bibitem{shape}   I. Bigi, M. Shifman, N. Uraltsev and A. Vainshtein,
Phys. Rev. Lett. 71, 496 (1993); Int. J. Mod. Phys. A 9, 2467
(1994); A. Manohar and M. Wise, Phys. Rev. D 49, 1310 (1994); M.
Neubert, Phys. Rev. D 49, 3392 and 4623 (1994); T. Mannel and M.
Neubert, Phys. Rev. D 50, 2037 (1994), U. Aglietti and G.
Ricciardi, Nucl. Phys. B 587, 363 (2000); U. Aglietti, Nucl. Phys.
B Proc. Suppl. 96, 453 (2001)

\bibitem{noi2}U.Aglietti, R.Sghedoni, L.Trentadue, hep-ph/0310360

\bibitem{rv}E.Remiddi, J.A.M.Vermaseren, Int. J. Mod. Phys. A15, 725 (2000)
\end{document}